\documentclass[a4paper]{jpconf}
\usepackage{graphicx}
\usepackage{epsfig}
\begin{document}

\title{Medium information from anisotropic flow and jet quenching in relativistic heavy ion collisions}
\author{Subrata Pal$^1$, Marcus Bleicher$^2$}
\address{$^1$Department of Nuclear and Atomic Physics, Tata Institute of
Fundamental Research, Homi Bhabha Road, Mumbai 400005, India} 
\address{$^2$Institut f\"ur Theoretische Physik, Goethe-Universit\"at and
Frankfurt Institute for Advanced Studies (FIAS), 60438 Frankfurt am Main, Germany}

\begin{abstract} 
Within a multiphase transport (AMPT) model, where the  initial conditions are obtained from 
the recently updated HIJING 2.0 model, the recent anisotropic flow and suppression data for 
charged hadrons in Pb+Pb collisions at the LHC energy of $\sqrt{s_{NN}} = 2.76$ TeV are 
explored to constrain the properties of the partonic medium formed. In contrast to RHIC,  
the measured centrality dependence of charged hadron multiplicity $dN_{\rm ch}/d\eta$ at LHC 
provides severe constraint to the largely uncertain gluon shadowing parameter $s_g$.
We find final-state parton scatterings reduce considerably hadron yield at midrapidity and 
enforces a smaller $s_g$ to be consistent with $dN_{\rm ch}/d\eta$ data at LHC.
With the parton shadowing so constrained, hadron production and flow over a wide transverse momenta 
range are investigated in AMPT. The model calculations for the elliptic and triangular flow are 
found to be in excellent agreement with the RHIC data, and predictions for the flow 
coefficients $v_n(p_T,{\rm cent})$ at LHC are given. The magnitude and pattern of suppression 
of the hadrons in AMPT are found consistent with the measurements at RHIC. However, the 
suppression is distinctly overpredicted in Pb+Pb collisions at the LHC energy. Reduction of 
the QCD coupling constant $\alpha_s$ by $\sim 30\%$ in the higher temperature plasma 
formed at LHC reproduces the measured hadron suppression.  

\end{abstract}

\section{Introduction}

Collisions of heavy nuclei at the Relativistic Heavy Ion Collider (RHIC) \cite{BRAHMS,PHOBOS,STAR,PHENIX} 
and recently at the Large Hadron Collider (LHC) \cite{ALICE,CMS} have created a matter consisting 
of deconfined but strongly coupled quarks and gluons (sQGP).
Evidence of this is provided by the hydrodynamic model analysis of elliptic flow data that 
requires an extremely small viscosity to entropy density ratio \cite{Paul,Song} and from the observed 
suppression of high-$p_T$ hadrons \cite{PHENIXsup,STARsup} in central collisions 
relative to both peripheral and nucleon-nucleon collision. The suppression has been well established as 
due to parton energy loss via medium induced gluon emission \cite{GLV}. The lost energy,
whose magnitude depends on the parton density, reappears as soft hadrons \cite{PalPratt,PalR}. 

As the QGP formed in central Pb+Pb collisions at $\sqrt{s_{NN}} = 2.76$ TeV is at a density of 2.4 larger 
and probes parton distribution at a smaller momentum fraction $x$ than at RHIC, analysis of the recent 
data for bulk hadron production \cite{ALICEnp,CMSnp}, anisotropic flow \cite{ALICEvn} and high-$p_T$ hadron 
suppression at LHC \cite{ALICEsup,CMSsup} may provide crucial insight into the nuclear medium 
properties of the hot and dense QCD matter.

%%%%%%%%%%%%%%%%%%%%%%%%%
\begin{figure}[h]
\hspace{0.5pc}%
\begin{minipage}{37pc} \hspace{4pc}%
\includegraphics[width=30pc]{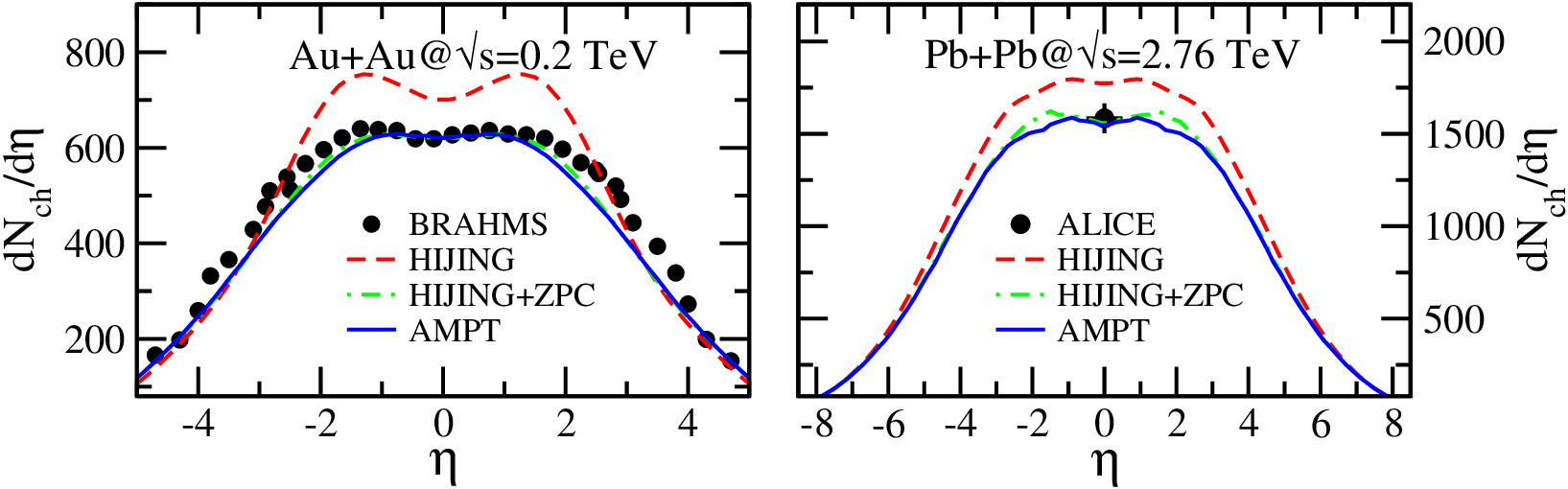}
\caption{\label{mulp}
$dN_{\rm ch}/d\eta$ distribution at RHIC (left panel) and LHC
(right panel) in 0-5$\%$ central collisions. The predictions from
HIJING (dashed line), HIJING+ZPC (dashed-dotted line), AMPT (solid line) are compared
to data (solid circles) from BRAHMS \cite{BRAHMSnp} and ALICE \cite{ALICEnp}. }
\end{minipage}
 \end{figure}
%%%%%%%%%%%%%%%%%%%%%%%%%

While perturbative QCD can address only hard scatterings, 
formation of sQGP and abundant soft particle production 
suggest a highly nonperturbative physics which is not yet well-established within QCD. 
Thus to explore medium effects on particle production from soft to the hard scattering regime 
relies on systematic inclusion of various stages of dynamical evolution of matter.

\section{The AMPT Model}

A MultiPhase Transport (AMPT) model \cite{AMPT} which combines the initial particle distribution 
from HIJING model \cite{HIJING} with subsequent parton-parton elastic scatterings via the
ZPC parton cascade model and final hadron transport via ART allows a systematic study of hadron production.
To investigate bulk and hard particle production, the AMPT model was modified to include the updated HIJING 
2.0 version. In the two-component HIJING model \cite{HIJING} for hadron production, nucleon-nucleon 
collision with transverse momentum $p_T$ transfer larger than a cut-off $p_0$ leads to 
jet production. Soft interactions with $p_T<p_0$ is characterized by an
effective cross section $\sigma_{\rm soft}$. In HIJING 2.0 \cite{HIJ2} the 
Duke-Owens parametrization \cite{Duke} of the parton distribution functions has been updated 
with the modern Gl\"uck-Reya-Vogt (GRV) parametrization \cite{GRV94}. Since the gluon 
distribution at small momentum fraction $x$ is much larger in GRV, instead of a fixed 
value for ($p_0, \sigma_{\rm soft}$) as in HIJING 1.0, an energy dependent cut-off 
for $p_0(\sqrt s)$ and  $\sigma_{\rm soft}(\sqrt s)$ are used to fit experimental data on 
total and inelastic cross sections in $p+p/{\bar p}$ collisions \cite{HIJ2}.

For the nuclear parton distribution function (PDF), HIJING employs the functional form
$f_a^A(x,Q^2) = A \: R_a^A(x,Q^2) \: f_a^N(x,Q^2)$, 
where $f_a^N$ is the PDF in a nucleon. The nuclear modification factor
of quarks and gluons ($a\equiv q,g$) in HIJING 2.0 parametrization are \cite{HIJ2}
\begin{eqnarray} \label{nmod}
R_q^A(x,b)&=& h(x) - s_q(b) \: (A^{1/3}-1)^{0.6} \: (1-3.5x^{0.5}) \exp(-x^2/0.01) ,  \nonumber\\
R_g^A(x,b) &=& h(x) - s_g(b) \: (A^{1/3}-1)^{0.6} \: (1-1.5x^{0.35}) \exp(-x^2/0.004) . 
\end{eqnarray}
Here $h(x) =  1 + 1.2\log^{1/6}\!\!A \ (x^3-1.2x^2+0.21x)$, the 
impact parameter dependence of shadowing is taken as $s_a(b)=(5s_a/3)(1-b^2/R_a^2)$, and 
$R_A \sim A^{1/3}$ is the nuclear size.  $s_q=0.1$ is fixed from deep inelastic scattering data
and $s_g$ shall be estimated from fits to the measured charged particle pseudorapidity density 
in heavy ion collisions. We have used here the string melting version
of the AMPT where parton recombination is employed for hadronization. 
At both RHIC and LHC energies, we consider the strong coupling 
constant $\alpha_s=0.33$ and screening mass $\mu =3.226$ fm$^{-1}$ \cite{XuKo} that 
correspond to parton-parton elastic scattering cross section of 
$\sigma \approx 9\pi\alpha_s^2/(2\mu^2) \approx 1.5$ mb.

%%%%%%%%%%%%%%%%%%%%%%%%%
\begin{figure}[h]
\hspace{1pc}%
\begin{minipage}{36pc} \hspace{1pc}%
\includegraphics[width=34pc]{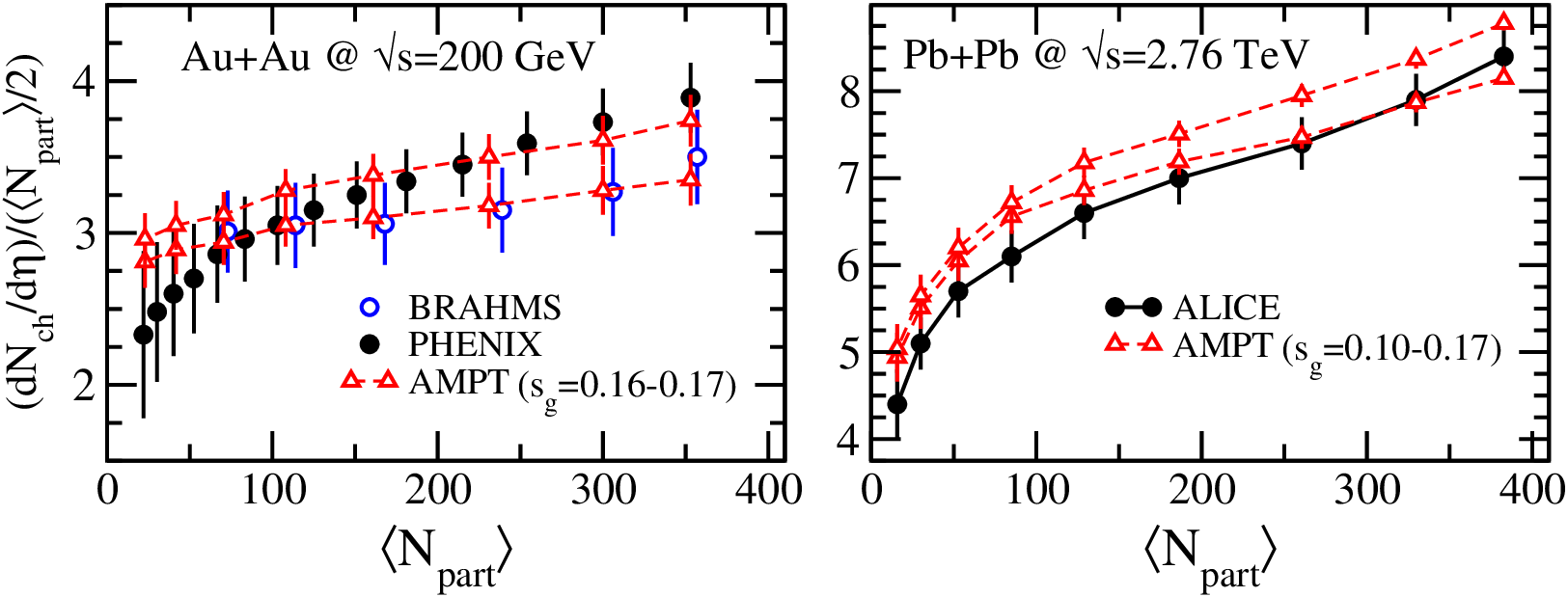} 
\caption{\label{npart}
$dN_{\rm ch}/d\eta$ at mid-rapidity per participant nucleon pair as a function of 
$\langle N_{\rm part}\rangle$.  The results are from AMPT calculations (triangles) with gluon 
shadowing $s_g$=0.10-0.17 at RHIC (left panel) and with $s_g=0.16-0.17$ at LHC (right panel) 
as compared with the data (circles) from BRAHMS \cite{BRAHMSnp} and PHENIX \cite{PHENIXnp} 
at RHIC and ALICE \cite{ALICEnp} at LHC.}
\end{minipage}
\end{figure}
%%%%%%%%%%%%%%%%%%%%%%%%%%%%%%%%%%%%%%

\section{Results and discussions}

\subsection{Particle yield distribution}

Figure \ref{mulp} shows the pseudorapidity distribution of charged hadrons, 
$dN_{\rm ch}/d\eta$, for central collisions in Au+Au at $\sqrt{s_{NN}}=200$ GeV and 
Pb+Pb at $\sqrt{s_{NN}} = 2.76$ TeV. The AMPT results are with gluon shadowing parameter
of $s_g=0.15$ (at RHIC) and $s_g=0.17$ (at LHC) that agrees well with 
the measured $dN_{\rm ch}/d\eta$ distribution from BRAHMS at 
RHIC, and the $dN_{\rm ch}/d\eta \: (|\eta|<0.5) = 1601\pm 60$ from ALICE at the LHC. 
The HIJING 2.0, without any final state interactions, predicts 
$dN_{\rm ch}/d\eta \: (|\eta|<0.5) = 706\pm 5$ and $1775 \pm 3$ at RHIC and LHC, respectively. 
Energy dissipation and redistribution via partonic scatterings in subsequent 
parton cascade (i.e. in HIJING plus ZPC) result in the reduction of charged particle yield
by $\sim 15\%$ at both RHIC and LHC inspite of $\sim 2.4$ times increase
in the partonic density at LHC \cite{PalBl}. Finally scatterings in the relatively 
dilute hadronic matter (i.e. AMPT) lead to only a small decrease of particle abundances.  

Figure \ref{npart} shows the charged particle pseudorapidity density per participant pair, 
$(dN_{\rm ch}/d\eta)/(\langle N_{\rm part} \rangle/2)$, as a function of centrality of collision 
characterized by average number of participating nucleons $\langle N_{\rm part} \rangle$. 
The AMPT calculations are performed with a range of gluon shadowing parameter of $s_g=0.10-0.17$ 
at the RHIC energy and with $s_g=0.16-0.17$ at the LHC.
This choice of $s_g$ leads to centrality dependence of charged particle multiplicity that 
agrees well within the experimental uncertainty of the  BRAHMS \cite{BRAHMSnp} 
and PHENIX \cite{PHENIXnp} data at RHIC. Due to abundant jet and minijet production in Pb+Pb collisions
at LHC, the ALICE multiplicity results are quite sensitive to nuclear distortions at
small $x$ and provide a much stringent constraint on the gluon shadowing of $s_g \simeq 0.17$.
We note that the estimated values of $s_g$ in AMPT are consistently smaller than the HIJING 2.0 
\cite{HIJ2} estimates of $s_g=0.17-0.22$ (RHIC) and $s_g=0.20-0.23$ (LHC) which indicate the 
importance of final state interactions in precise determination of the nuclear shadowing of partons.

%%%%%%%%%%%%%%%%%%%%%%%%%%%%%%%%%%%%
\begin{figure}[h]
\begin{minipage}{38pc}
\includegraphics[width=11pc]{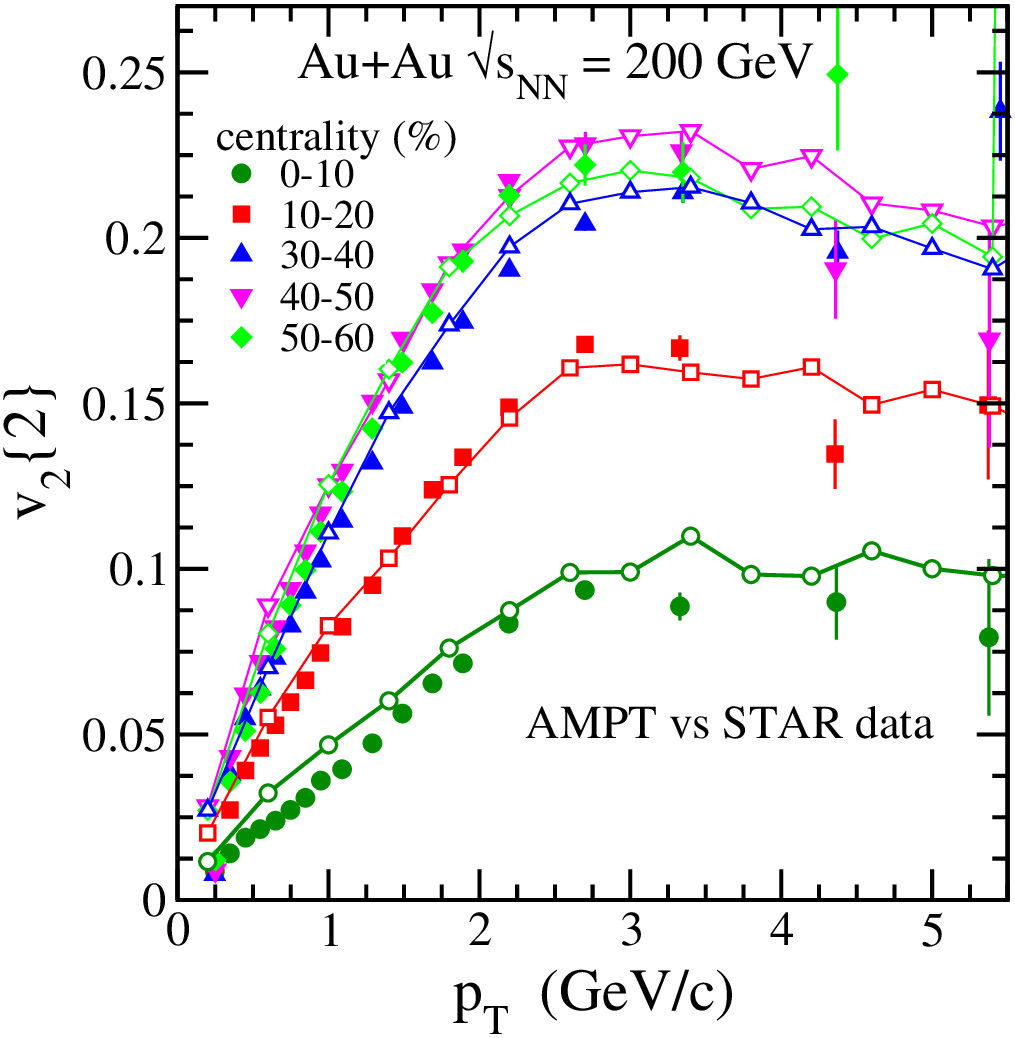} \hspace{1.5pc}%
\includegraphics[width=11.7pc]{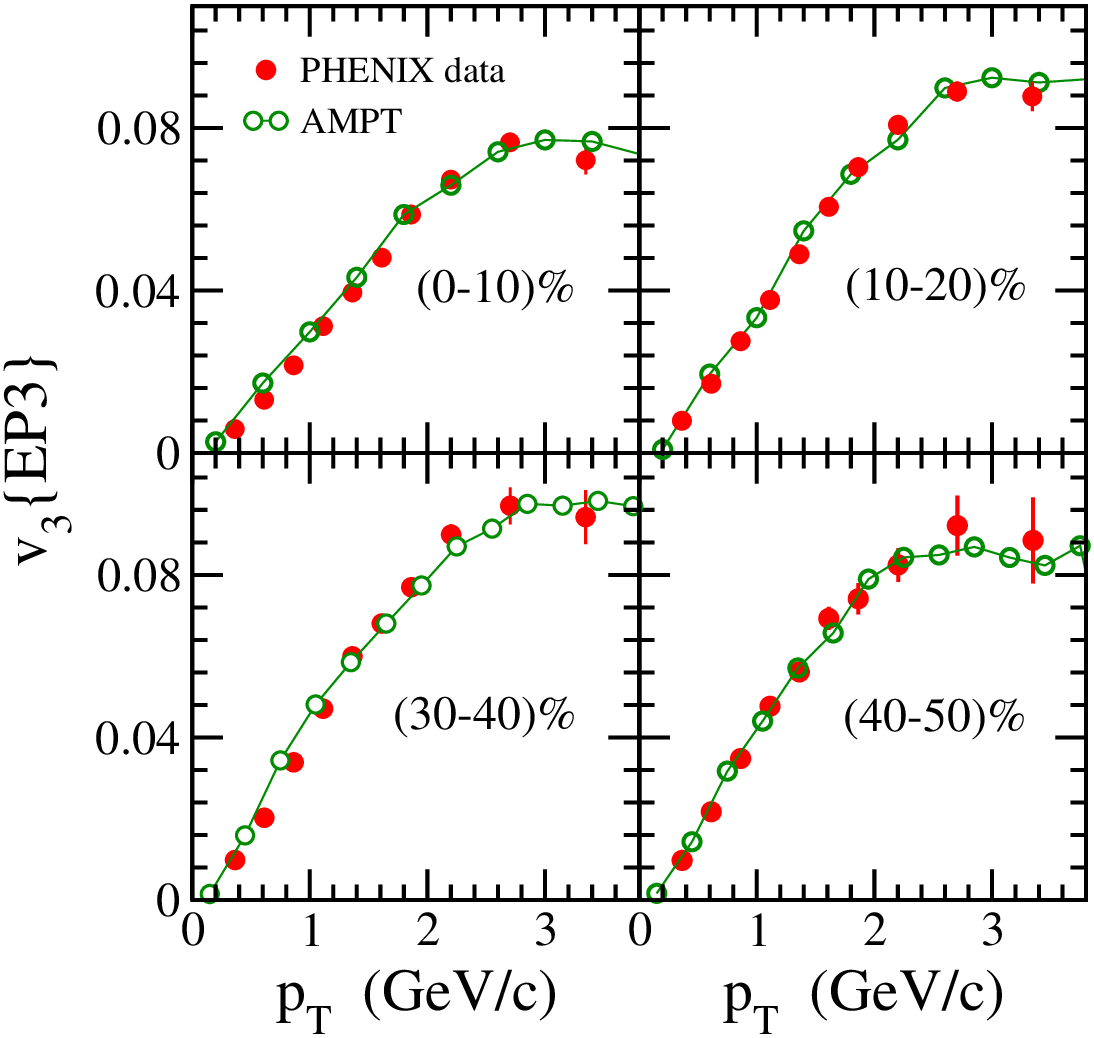} \hspace{1.5pc}%
\includegraphics[width=11pc]{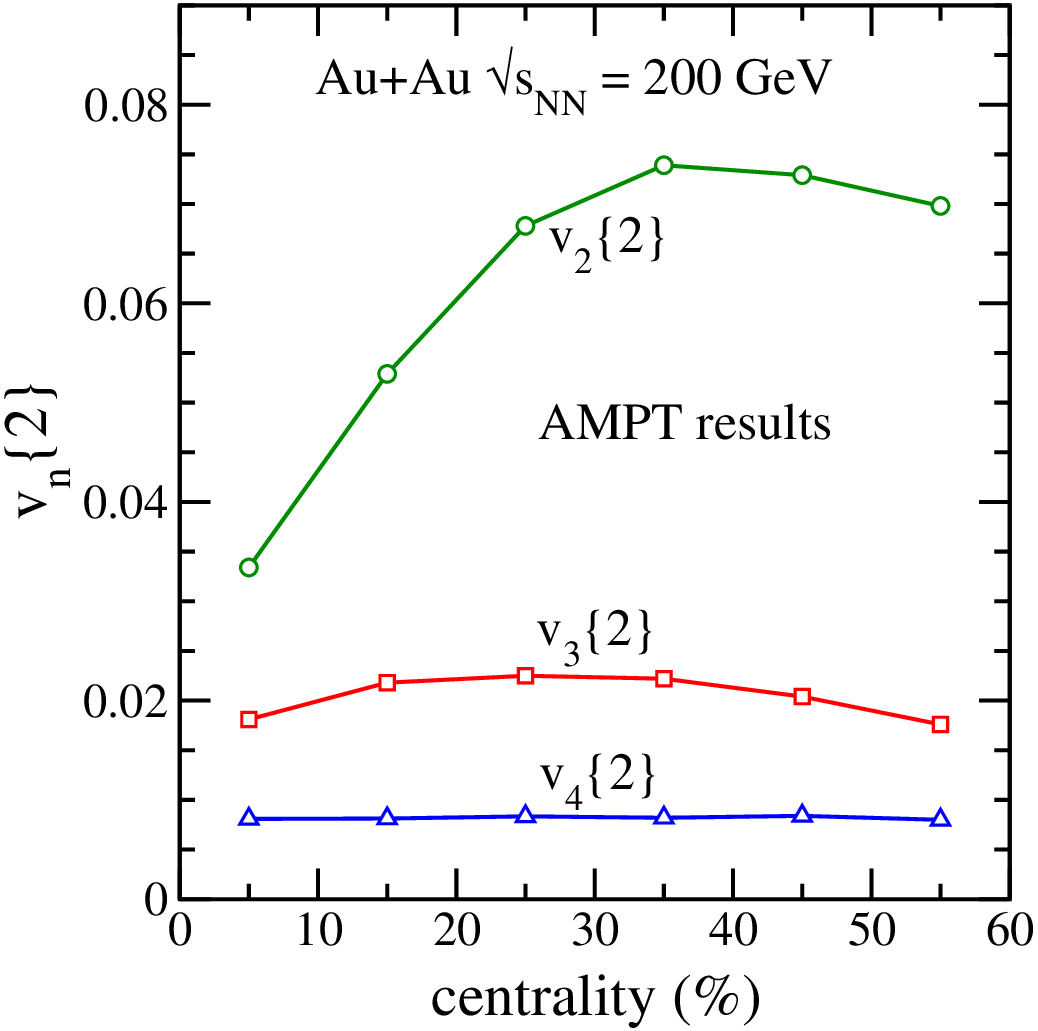}
%\caption{\label{v2}$v_2$ vs. $p_T$ at RHIC (top panel) and LHC (bottom panel).}
\end{minipage} 
\end{figure}
%%%
\begin{figure}[h]
\begin{minipage}{38pc}
\includegraphics[width=11pc]{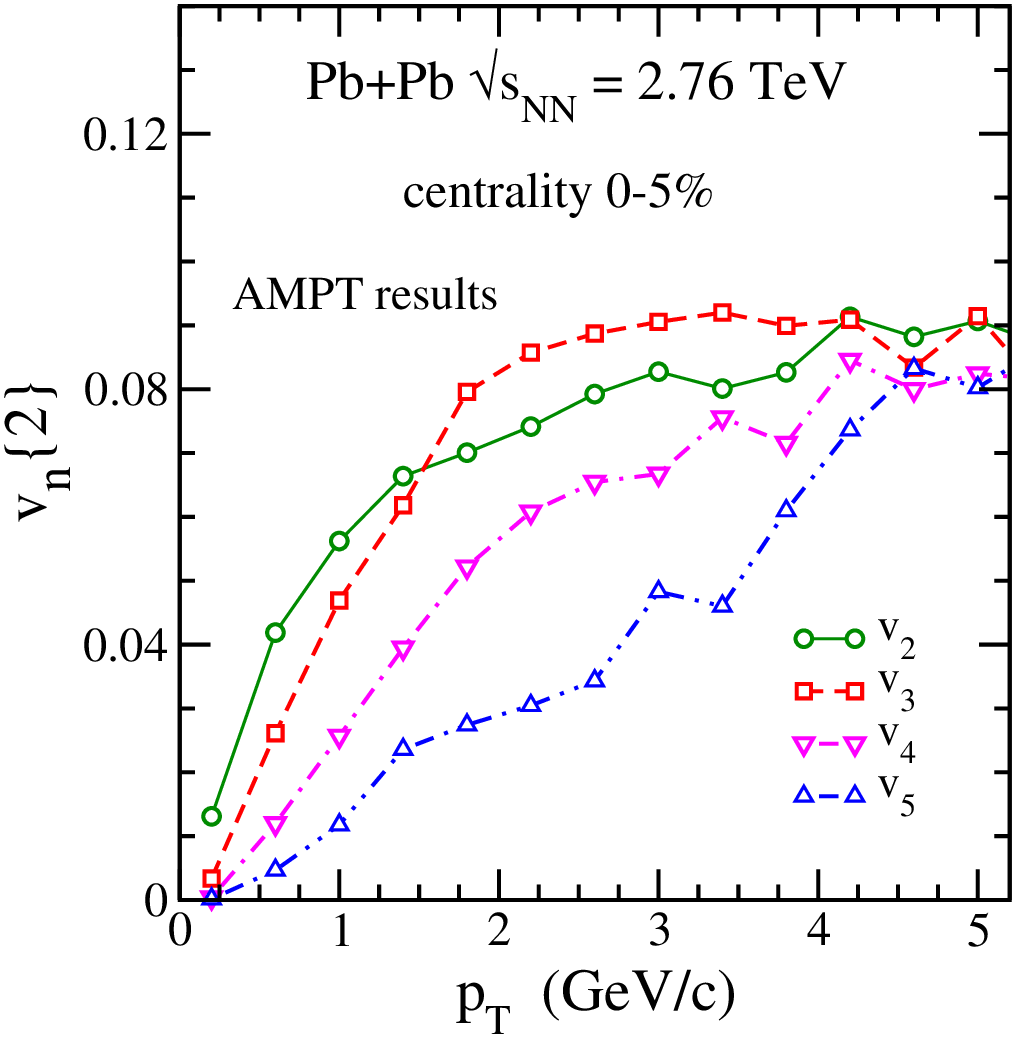}\hspace{2.2pc}%
\includegraphics[width=11pc]{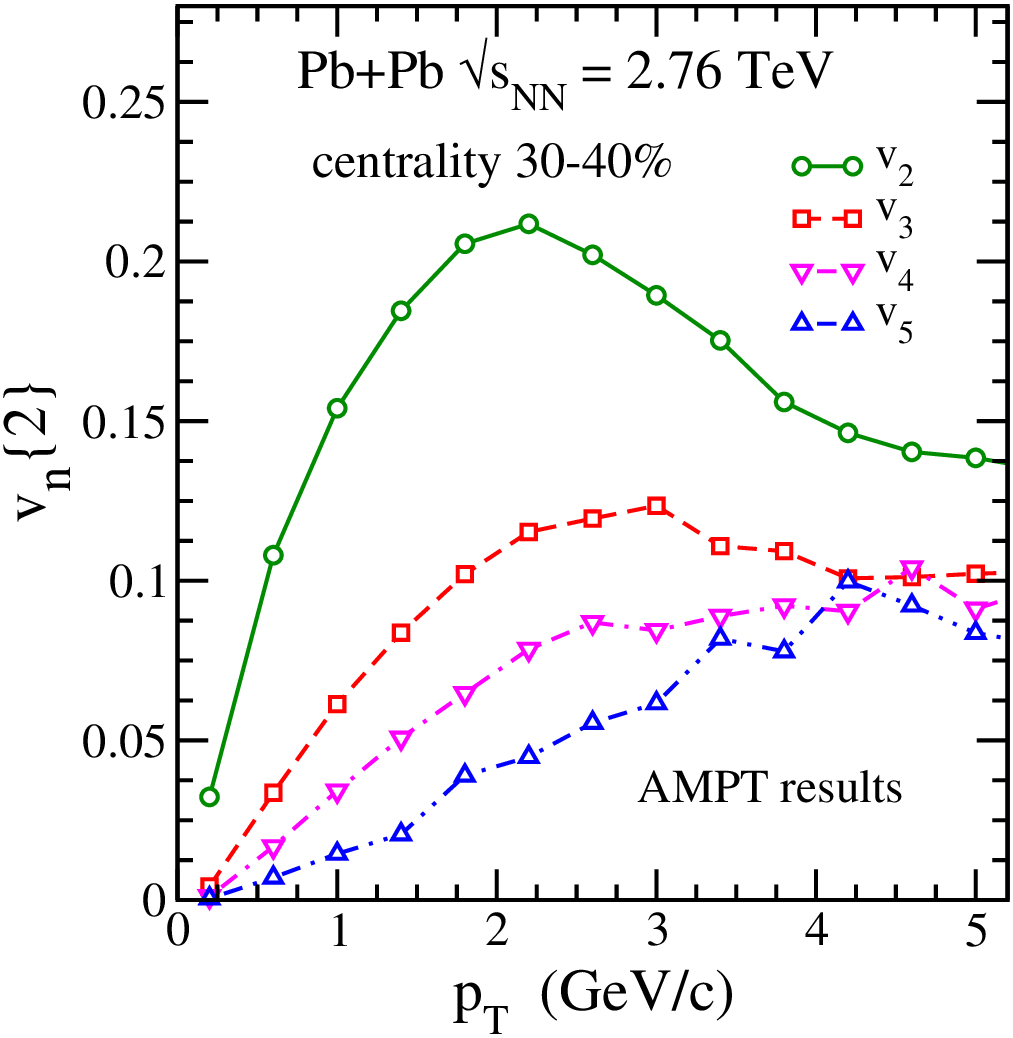}\hspace{2.2pc}%
\includegraphics[width=11pc]{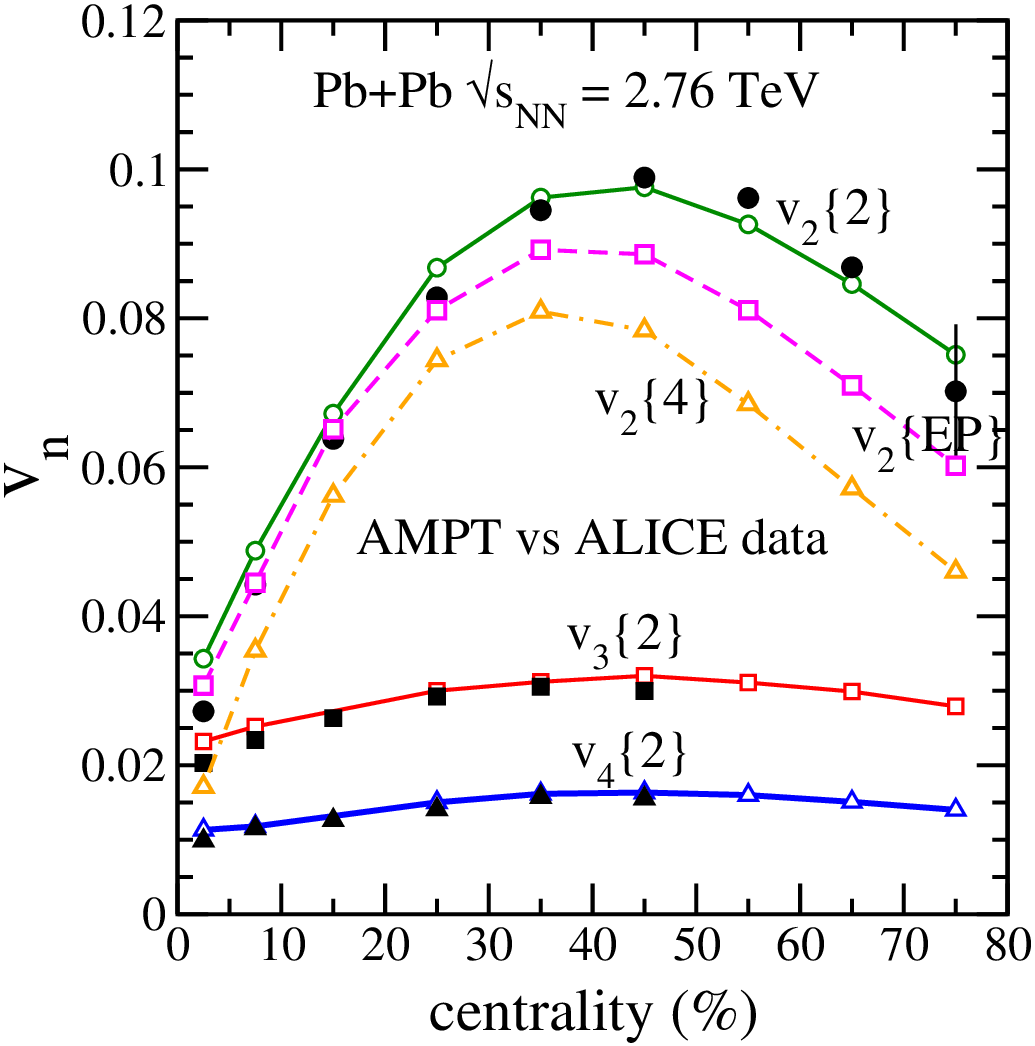}
\caption{\label{vnal}Top panels: $p_T$ dependence of $v_2$ (left panel), $v_3$ (middle panel) at
various centralities, and centrality dependence of $v_n$ ($n=2,3,4$) (right panel) at RHIC.
The solid symbols are STAR \cite{STARvn} and PHENIX \cite{PHENIXv3} data.
Bottom panels: $p_T$ dependence of $v_n$ at centralities of 0-5$\%$ (left panel), 
30-40$\%$ (middle panel), and centrality dependence of $v_n$ (right panel) at LHC.
The solid symbols are ALICE data \cite{ALICEvn}.} 
\end{minipage}
\end{figure}
%%%%%%%%%%%%%%%%%%%%%%%%%%%%%%%%%%%%

\subsection{Anisotropic Flow}

Partonic interactions within the initial spatially asymmetric collision zone drives uneven 
pressure gradients that manifests in anisotropic emission of particles leading to collective flow. 
The magnitude of this flow is characterized \cite{Borghini,Alver} by the Fourier coefficients 
$v_n = \langle \cos(n[\phi - \Psi_n])\rangle$ as elliptic ($v_2$), triangular ($v_3$) and quadrangular
($v_4$) flow estimated with respect to their participant event planes $\Psi_n$. Characterization 
can also be made via the correlations between the $k$-particles with azimuthal angles 
$\phi_1,\ldots,\phi_k$ as $v\{n_1,\ldots,n_k\}=\langle \cos(n_1\phi_1+\cdots+n_k\phi_k)\rangle$, 
where $n_1,\ldots,n_k$ are integers and the average is taken over all charged particles and events.
This allows to construct the 2- and 4-particle correlations, $v_n\{2\} \equiv \sqrt{v\{n,-n\}}$ 
and $v_n\{4\}$, respectively.

In Fig. \ref{vnal} (top panel),  the transverse momentum dependence of elliptic flow $v_2\{2\}(p_T)$ for
charged hadrons at 
midrapidity is displayed at the RHIC energy at various centralities. It is seen that over a large $p_T$ 
range the AMPT model is consistent with the STAR data \cite{STARvn}. Spatial fluctuations of participating 
nucleons lead to $v_3$ of similar magnitude as $v_2$ for most central collisions in AMPT.
While the triangularity in the collision geometry, and thus $v_3$, is nearly insensitive to increase in 
collision peripherality (impact parameter), the $v_2$ increases as the initial spatial eccentricity becomes 
more pronounced.  We find AMPT also reproduces the measured 
$v_3\{{\rm EP}\}(p_T)$ (top-middle panel) as well as $v_4\{{\rm EP}\}(p_T)$ 
(not shown here) in the event-plane method at various centralities with the same cross section 
of $\sigma \approx 1.5$ mb and shadowing $s_g=0.15$ at the RHIC energy as used/estimated above. 
The AMPT model predictions for the centrality dependence of $v_n\{2\}$ ($n=2,3,4$) with momenta 
$0.2 <p_T < 8$ GeV/c are presented in Fig. \ref{vnal} 
(top-right panel) at the RHIC energy. With collision centrality, the $p_T$-averaged $v_2$ 
shows substantial variation, the changes are small in $v_3$ and negligible in $v_4$. 

For central Pb+Pb collisions at the LHC energy, the $p_T$ dependence of $v_n\{2\}$ as seen in
Fig. \ref{vnal} (bottom panel) exhibits similar magnitude and pattern as at RHIC energy. 
The enhanced parton density at the LHC energy that is expected to produce larger $v_2$ appears to be 
compensated by faster expansion of the evolving medium due to larger flow. The centrality dependence 
of $v_n\{2\}$ ($n=2,3,4$) is found to agree quite well with the ALICE data \cite{ALICEvn}.
As expected the smaller nonflow effects result in successively reduced elliptic flow in the 
event-plane $v_2\{{\rm EP}\}$ and 4-particle correlation $v_2\{4\}$ methods.

%%%%%%%%%%%%%%%%%%%%%%%
\begin{figure}[h]
\hspace{1pc}% 
\begin{minipage}{36pc} 
\includegraphics[width=32pc]{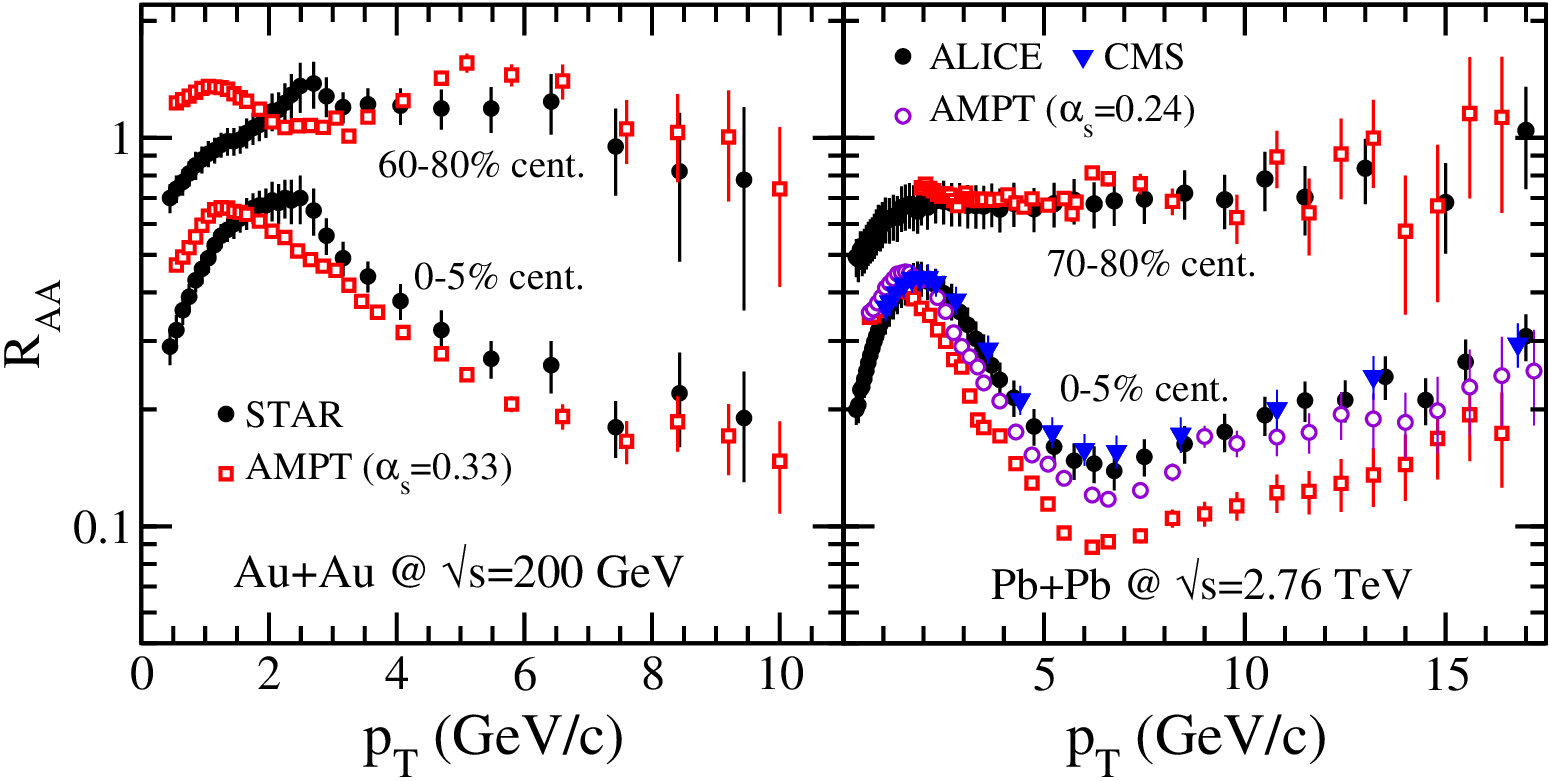}
\caption{\label{RAA}
$R_{AA}$ for charged hadrons as a function of $p_T$ in central and peripheral collisions 
for Au+Au at RHIC (left panel) and Pb+Pb at LHC (right panel). The data are from
STAR \cite{STARsup} at RHIC and from ALICE \cite{ALICEsup} and CMS \cite{CMSsup} at LHC. 
The AMPT results are with strong coupling constant $\alpha_s=0.33$ at RHIC and LHC and 
with $\alpha_s=0.24$ for central collisions at LHC}
\end{minipage}
\end{figure}
%%%%%%%%%%%%%%%%%%%%%%%%

\subsection{High-$p_T$ charged hadron suppression}

The study of bulk hadron production in conjunction with that for hadron spectra
provide crucial information on the parton-medium interactions where high-$p_T$ partons
suffer energy loss which in turn produce soft hadrons. 
The suppression of hadrons at high $p_T$ due to medium effects in heavy ion 
collisions is quantified by the nuclear modification factor
\begin{equation} \label{RAAeq}
R_{AA}(p_T) = \frac{d^2N^{AA}/d\eta \: dp_T} 
{\langle N_{\rm coll}\rangle \ d^2N^{pp}/d\eta \: dp_T} 
\end{equation}
which refers to the ratio of particle yield in heavy ions ($A+A$) to that in $p+p$ reference 
spectra, scaled by the total number of binary nucleon-nucleon ($NN$) collisions  
$\langle N_{\rm coll}\rangle = \langle T_{AA}\rangle \sigma^{NN}_{\rm inel}$. 
In absence of initial and final state nuclear medium effects $R_{AA}(p_T)=1$ by construction.
The nuclear thickness function $\langle T_{AA}\rangle$ and the inelastic $NN$ cross section
$\sigma^{NN}_{\rm inel}$ are calculated within the HIJING 2.0 that employs Glauber
model for the distribution of initial nucleons with a Woods-Saxon nuclear density.

The nuclear modification factor $R_{AA}$ for charged hadrons is shown in Fig. \ref{RAA}. 
For central collisions at RHIC and LHC, $R_{AA}(p_T) < 1$ which suggests
appreciable suppression of particles relative to $NN$ reference. With the nuclear shadowing parameter 
$s_g=0.15$ constrained from $dN_{\rm ch}/d\eta$ data in Au+Au collisions, the AMPT results describe the 
magnitude and pattern of the RHIC suppression data \cite{STARsup}. The success of AMPT at RHIC suggests 
that the initial state shadowing, the final state scattering and the parton energy loss
is consistent with the formation and evolution of the medium at the RHIC energy.

In peripheral Pb+Pb collisions at LHC, the $R_{AA}$ for $h^\pm$
is nearly constant at about 0.7 over a large $p_T$ range. Here, 
the QGP even if formed, should have a small volume and short lifetime. 
In central Pb+Pb collisions at LHC, the rise and fall pattern exhibited by $R_{AA}$ up to 
$p_T \sim 6$ GeV/c (in data and model) is similar to that at RHIC. 
The subsequent rise of $R_{AA}$ at LHC is mainly due to the harder unquenched pQCD jet spectra. 
However, in contrast to ALICE and CMS data, the AMPT calculations show even more pronounced suppression 
at $p_T>2$ GeV/c. This suggests that the LHC medium with a factor of 2.4 increase in density over RHIC 
is, in fact, more transparent \cite{PalBl,Horowitz}.

It may be noted that at both RHIC and LHC, the $R_{AA}$ was evaluated in the AMPT with the same value 
for the QCD coupling constant $\alpha_s=0.33$ and screening mass $\mu =3.226$ fm$^{-1}$.
As the screening mass depends on temperature as $\mu = gT = \sqrt{4\pi\alpha_s}T$ \cite{Blaizot},
the parton-parton elastic scattering cross section used in AMPT becomes 
$\sigma \approx 9\pi\alpha_s^2/(2\mu^2) \approx 9\alpha_s/(8T^2)$. Using the scaling relation
between the initial entropy density $s_i$ and the final particle multiplicity
\cite{Gyulassy,PPs} $s_i \approx (7.85/\tau_i A_\perp) \: dN_{\rm ch}/dy$,
and considering a QGP of massless gas of light quarks and antiquarks so that 
$s_i \approx 4 \epsilon_i /(3T_i)$ (with energy density 
$\epsilon_i \approx (21/30)\pi^2 T_i^4$), allows one to estimate the initial temperature 
$T_i$ and thereby the parton scattering cross section $\sigma$ from 
the measured particle yield. In $0-5\%$ central Au+Au and Pb+Pb collisions at   
$\sqrt{s_{NN}} = 0.2$ and 2.76 TeV, the measured $dN_{\rm ch}/dy \approx 687$ \cite{PHENIXnp} 
and 1601 \cite{ALICEnp} result in $T_i \approx 320$ and 436 MeV respectively, at a proper 
time $\tau_i=1$ fm/c. With the above choice of $\alpha_s=0.33$, the estimated
$\sigma \approx 9\alpha_s/(8T_i^2) \approx 1.4$ mb at RHIC turns out to be similar to the 
value employed in AMPT that reproduces the RHIC suppression data shown in Fig. \ref{RAA}. 
On the other hand, the higher $T_i$ at LHC leads to $\sigma \approx 0.76$ mb. Alternatively, 
if $\mu$ is set constant at 3.226 fm$^{-1}$ from RHIC to LHC, this smaller $\sigma$ then gives 
$\alpha_s \approx 0.24$ at LHC. With this reduced $\alpha_s$, we find from Fig. \ref{RAA} 
the AMPT results for $R_{AA}$ (open circles) in central Pb+Pb collisions at LHC is in 
good agreement with the suppression data. This is a clear indication of thermal 
suppression of the QCD coupling constant at the higher temperature reached at LHC.

\section{Conclusions}

In summary, within the AMPT model that is updated with the HIJING 2.0 initial conditions 
for parton distributions, we study the properties of the medium formed in heavy ion collisions 
at RHIC and LHC energies. We find energy redistribution via parton scatterings reduces the final 
hadron multiplicity and thereby enforces a smaller gluon shadowing for the hadron yield
to be consistent with the data. The magnitude and trend of the flow coefficients, $v_2$, $v_3$, 
$v_4$ as a function of $p_T$ and at various centralities are in excellent agreement with the 
measurements at RHIC and LHC. The parton energy loss in AMPT is found to describe the charged hadron 
suppression over a large $p_T$ range at RHIC. However, using the same QCD couplings $\alpha_s$ 
and parton scattering cross section from RHIC to LHC, jet quenching in AMPT is distinctively 
overpredicted relative to the measurements at central Pb+Pb collision at LHC.
A reduction of $\alpha_s$ by $\sim 30\%$, consistent with the higher temperature of the plasma 
formed at LHC, agrees with the measured suppression.

\ack
We thank Andre Yoon for providing the experimental CMS suppression data. SP acknowledges
support from the Alexander von Humboldt Foundation and kind hospitality at FIAS where
part of the work was completed.

\end{document}